\documentclass{raa}            
\usepackage{graphicx,times}
\usepackage{natbib}
\usepackage{amssymb,amsmath}
\bibpunct{(}{)}{;}{a}{}{,}

\begin{document}
\title{Orbit and spin evolution of the synchronous binary stars on the main
sequence phase }
 \volnopage{ {\bf 2012} Vol.\ {\bf 12} No. {\bf XX}, 000--000}
   \setcounter{page}{1}

\author{Lin-Sen Li}

\institute{
School of physics, Northeast Normal University, Changchun 130024, China; {\it lils653@nenu.edu.cn}\\
\vs \no
   {\small Received 2010 August 20; accepted 2012 May 17}
}

\abstract{The sets of the synchronous equations are
derived from the sets of non-synchronous equations The analytical
solutions are given by solving the set of differential equations.
The results of the evolutionary tendency of the orbit-spin are that
the semi-major axis shrinks gradually with time: the orbital
eccentricity dereacses gradually with time until the orbital
circularization; the orbital period shortens gradually with time and
the rotational angular velocity of primary component speed up with
time gradually before the orbit-rotation achieved the
circularization The theoretical results are applied to evolution of
the orbit and spin of synchronous binary stars Algol A, B on the
main sequence phase The circularization time and life time (age) and
the evolutional numerical solutions of orbit and spin when
circularization time are estimeted for Algol A, B. The results are
discussed and concluded. }

   \authorrunning{L. S. Li }            
   \titlerunning{Orbit and Spin Evolution of the Synchronous Binary Stars}  
   \maketitle

\section{Introduction}

The tidal friction plays an important role in the evolution of the
orbit and spin of close binary system. Earliest, the author
researching this topic is \cite{Zahn+1965, Zahn+1966a, Zahn+1966b,
Zahn+1966c, Zahn+1975}. \cite{Alexander+1973} firstly studied the
dynamical problem of the tidal friction in close binary system by
using the method employed by \cite{Darwin+1879}. Later on,
\cite{Hut+1980, Hut+1981} generalized the method given by
\cite{Alexander+1973}. He studied the stability of tidal equilibrium
and tidal evolution in close binary system by using the method of
energy and angular momentum. But their research dealt with a few
synchronization. The sequential research for the synchronization of
rotation are given by \cite{Zahn+1977, Zahn+1978}.
\cite{Rajamohan+Venkatakrishnan+1981} ever studied synchronization
in binary stars. \cite{Giuricin+etal+1984a} researched
synchronization in eclipsing binary stars and
\cite{Giuricin+etal+1984b} also researched synchronization in
early-type spectroscopic binary stars, \cite{Zahn+Bouchet+1989b}
studied mainly the orbital circularization of late-type binary stars
on the pre-main sequence and the theoretical results are given based
on \cite{Zahn+1989a}. \cite{Pan+1996} calculated the circularization
time scale by using the two mechanisms: one is the equilibrium tidal
mechanism given by \cite{Zahn+1977}, another is purely hydrodynamic
mechanissm given by Tassoul~(\citeyear{Tassoul+1987}).
\cite{Keppens+etal+2000} studied the rotational evolution of binary
stars system: synchronization and circularization.
\cite{Huang+Zeng+2000} also researched evolution of non-synchronized
binary stars with 9 solar mass and 6 solar mass.
\cite{Meibom+etal+2005} studied obseration tidal synchronization in
detached solar-type binary stars and \cite{Meibom+etal+2006} also
researched an observational study of tidal synchronization in
solar--type binary starss in open clusters M35 and M34. Although the
author \cite{Li+1998, Li+2004, Li+2009} studied some methods for
judging the synchronization of rotation of binary stars, but he does
not studied the evolution of orbit-rotation of synchronous binary
stars. In the present paper the author examined the evolutional
tendency of orbit and spin of synchronous binary stars on the main
sequence phase.

\section{The evolutional equations of synchronous binary stars due to the
tidal friction on the main sequence}

The secular evolutional equations of the semi-major axis, $a$, eccentricity,
$e$, and the rotational angular velocity, $\Omega $, due to tidal frication
for non-synchronous binary stars are given by \cite{Zahn+1989a}.
\begin{eqnarray}
\label{eq1}
\frac{1}{a}\frac{da}{dt}&=&-\frac{12}{t_f }q(1+q)\Big(\frac{R}{a}\Big)^8\left\{
{\lambda ^{22}\Big(1-\frac{\Omega }{\omega }\Big)} \right.\nonumber\\
&+&e^2\Big[\frac{3}{8}\lambda ^{10}+\frac{1}{16}\lambda ^{12}\Big(1-2\frac{\Omega
}{\omega }\Big)-5\lambda ^{22}\Big(1-\frac{\Omega }{\omega }\Big)+\frac{147}{16}\lambda
^{32}\left. {\Big(3-2\frac{\Omega }{\omega }\Big)\Big]} \right\} ,
\end{eqnarray}
\begin{eqnarray}
\label{eq2}
\frac{1}{e}\frac{de}{dt}&=&-\frac{3}{t_f
}q(1+q)\Big(\frac{R}{a}\Big)^8\Big[\frac{3}{4}\lambda ^{10}-\frac{1}{8}\lambda
^{12}\Big(1-2\frac{\Omega }{\omega }\Big)\nonumber\\
&-&\lambda ^{22}\Big(1-\frac{\Omega }{\omega }\Big)+\frac{49}{8}\lambda
^{32}\Big(3-2\frac{\Omega }{\omega }\Big)\Big],
\end{eqnarray}
\begin{eqnarray}
\label{eq3}
\frac{d}{dt}(I\Omega)&=&\frac{6}{t_f}q^2MR^2\Big(\frac{R}{a}\Big)^6\left\{ {\lambda
^{22}(\omega -\Omega)+e^2\Big[\frac{1}{8}\lambda ^{12}(\omega -2\Omega))}
\right.\nonumber\\
&-&5\lambda ^{22}(\omega -\Omega)+\left. {\frac{49}{8}\lambda ^{32}(3\omega
-2\Omega)\Big]} \right\}.
\end{eqnarray}
Where $M$ and $R$ denote the mass and the radius of primary star,
$q=\frac{{M}'}{M}$, $M'$ denotes the mass of the secondary star, $\omega$
denotes the orbital angular velocity (mean motion), $I$ denotes momentum of
inertia, the convective friction time and $\lambda ^{lm}$ are given by
(\citealt{Zahn+Bouchet+1989b})
\[
t_f =[\frac{MR^2}{L}]^{1/3},
\quad
\lambda ^{lm}=\lambda _2 [{2\pi } \mathord{\left/ {\vphantom {{2\pi }
{1l\omega -m\Omega 1}}} \right. \kern-\nulldelimiterspace} {1l\omega
-m\Omega 1}].
\]
Here $L$ denotes the luminosity of the primary star.

Next, one derives the evolutional equations of synchronous binary
stars. \cite{Zahn+Bouchet+1989b} point out that when the two components rotate
with the orbital motion in synchronism: $\left| {l\omega -m\Omega }
\right|=\omega $, then, all tidal coefficients are identical $\lambda
^{lm}=\lambda $ except for $\lambda ^{22}$. Hence in Equations~(\ref{eq1}) -- (\ref{eq3})
$\lambda ^{11}=\lambda ^{10}=\lambda ^{12}=\lambda ^{32}=\lambda $, $\lambda
^{22}\ne \lambda $. When we consider the two components rotate in
synchronism, $\Omega =\omega $ or $\frac{\Omega }{\omega }=1$. Substituting
these conditions into the Equations~(\ref{eq1}) -- (\ref{eq3}), the secular Equations~(\ref{eq1}) --
(\ref{eq3}) are reduced to the following simplified secular synchronized equations
\begin{equation}
\label{eq4}
\frac{1}{a}\frac{da}{dt}=-114q(1+q)\frac{\lambda }{t_f
}e^2\Big(\frac{R}{a}\Big)^8,
\end{equation}
\begin{equation}
\label{eq5}
\frac{1}{e}\frac{de}{dt}=-21q(1+q)\frac{\lambda }{t_f }\Big(\frac{R}{a}\Big)^8,
\end{equation}
\begin{equation}
\label{eq6}
\frac{1}{\Omega }\frac{d\Omega }{dt}=36q^2\Big(\frac{MR^2}{I}\Big)\frac{\lambda
}{t_f }e^2\Big(\frac{R}{a}\Big)^6=36q^2\Big(\frac{M}{I}\Big)\frac{\lambda }{t_f
}e^2a^2\Big(\frac{R}{a}\Big)^8.
\end{equation}
We may also write down a supplementary secular equations according to
keplerian third law
\begin{equation}
\label{eq7}
\frac{1}{\omega }\frac{d\omega }{dt}=-\frac{3}{2}\frac{1}{a}\frac{da}{dt},
\end{equation}
\begin{equation}
\label{eq8}
\frac{1}{P_{\rm orb} }\frac{dP_{\rm orb}
}{dt}=\frac{3}{2}\frac{1}{a}\frac{da}{dt},
\end{equation}
\begin{equation}
\label{eq9}
\frac{1}{P_{\rm Rot} }\frac{dP_{\rm Rot} }{dt}=-\frac{1}{\Omega }\frac{d\Omega
}{dt}.
\end{equation}
Where $P_{\rm orb}$ denotes the orbital period and $P_{\rm Rot}$ denotes the
rotational period

Substituting the Equation~(\ref{eq5}) for $de/dt$ into the following equation, we get
the circurization scale time
\begin{equation}
\label{eqo10}
t_{\rm cir} =\frac{e}{\frac{de}{dt}}=\frac{t_f }{21q(1+q)\lambda
}\Big(\frac{a}{R}\Big)^8.
\end{equation}
In the following one uses the analytical method to solve the
evolutional Equations~(\ref{eq4})--(\ref{eq9}) with the eccentricity $e$ as an independent variable

This paper considers the evolutional tendency of the orbit-spin of
the synchronous binaries before the orbital circularization on the phase of
the main sequence. We may assume that the radius, $R$, of the primary star may
be regarded as no variation, i.e, $R$ is a constant on the phase of the main
sequence star, but their separation or semi-major axis is variable due to
the action of the tidal friction.

Combining the Equation~(\ref{eq4}) with (\ref{eq5}), we obtain the differential equation
\begin{equation}
\label{eq10}
\frac{1}{a}\frac{da}{de}=\frac{38}{7}e.
\end{equation}
Integrating this equation, we get
\begin{equation}
\label{eq11}
a=a_0 \exp [\frac{19}{7}(e^2-e_0 ^2)].
\end{equation}
Substituting the Equation~(\ref{eq11}) into the Equation~(\ref{eq5}), we obtain the
equation:
\begin{equation}
\label{eq12}
\frac{1}{e}\frac{de}{dt}=-21q(1+q)\frac{\lambda }{t_f }(\frac{R}{a_0
})^8\exp [-\frac{152}{7}(e^2-e_0 ^2)].
\end{equation}
By letting $c = 152/7$, the differential Equation~(\ref{eq12}) can be written
\begin{equation*}
\frac{1}{e}\exp c(e^2-e_0 ^2)de=\exp (-ce_0 ^2)\frac{\exp
(ce^2)}{e}de=-21q(1+q)\frac{\lambda }{t_f }\big(\frac{R}{a_0}\big)^8 dt.
\end{equation*}
Using the expansion of series
\[
\exp (ce^2)=1+ce^2+\frac{1}{2}c^2e^4+\frac{1}{3}c^3e^6+.......
\]
Integrating the differential equation above, yields
\[
\exp (-ce_0 ^2)[\ln (e)+\frac{1}{2}ce^2+\frac{1}{8}c^2e^4+......]^e_{e_0}=-21q(1+q)\frac{\lambda }{t_f }(\frac{R}{a_0 })^8(t-t_0).
\]
We obtain time scale in terms of eccentricity, $e$, by neglecting the term
with $e^4$
\begin{equation}
\label{eqo14}
t-t_0 =-\frac{\ln (\frac{e}{e_0 })+\frac{1}{2}c(e^2-e_0 ^2)}{21q(1+q)Q}.
\end{equation}
Here
\begin{equation}
\label{eqo15}
Q=\frac{\lambda }{t_f }\big(\frac{R}{a_0}\big)^8\exp \big(\frac{152}{7}e_0 ^2\big).
\end{equation}

Combining the Equation~(\ref{eq4}) with (\ref{eq6}), we get the equation
\begin{equation*}
ada=-\frac{57}{18}\frac{(1+q)}{q}\big(\frac{I}{M}\big)\frac{d\Omega }{\Omega }.
\end{equation*}
Integrating this equation, we obtain
\begin{equation}
\label{eq13} \Omega =\Omega _0 \exp
\Big[-\frac{9}{57}\big(\frac{q}{1+q}\big)\big(\frac{M}{I}\big)(a^2-a_0
^2)\Big].
\end{equation}
The from Equation~(\ref{eq11}) {$a=a_0\exp\big[\frac{19}{7}(e^2-e_0^2)\big]$, so $a^2=a_0^2\exp\big[2\times\frac{19}{7}(e^2-e_0^2)\big]$, hence}
\begin{equation}
\label{eq14} a^2-a_0 ^2=a_0 ^2 \left\{\exp \big[\frac{38}{7}(e^2-e_0
^2)\big]-1 \right\}.
\end{equation}
We obtain the variation of the angular velocity of primary in terms of
eccentricity, $e$
\begin{equation}
\label{eq15} \Omega =\Omega _0 \exp \left\{
{-\frac{9}{57}\big(\frac{q}{1+q}\big)\frac{M}{I}a_0 ^2\{\exp
\big[\frac{38}{7}(a^2-a_0 ^2)\big]-1\}} \right\}.
\end{equation}
The integrations of the Equations~(\ref{eq7}) -- (\ref{eq9}) can be obtained
\begin{equation}
\label{eq16} \omega =\omega _0 \exp [-\frac{57}{14}(e^2-e_0 ^2)],
\end{equation}
\begin{equation}
\label{eq17} P_{\rm orb} =(P_0)_{\rm orb} \exp
[+\frac{57}{14}(e^2-e^2_0)],
\end{equation}
\begin{equation}
P_{\rm Rot} =(P_0)_{\rm Rot} \exp \left\{
{+\frac{9}{57}\big(\frac{q}{1+q}\big)\frac{M}{I}a_0 ^2\{\exp
\big[\frac{38}{7}(a^2-a_0 ^2)\big]-1\}} \right\}.
\end{equation}

{Next, one gives the analytical solutions of the secular evolutional
equations with time $t$ as an independent variable on the main
sequence.}

For small eccentricity, $e$, such as Algol A, B, $e = 0.015$, so
$\frac{1}{2}ce^2=0.0024$, the second term of the right hand of formula~(\ref{eqo14}
) may be neglected., and then, we can obtain the eccentricity decreases with
time from the formula~(\ref{eqo14})
\begin{equation}
\label{eq18}
e=e_0 \exp [-21q(1+q)Q(t-t_0)].
\end{equation}
\begin{equation}
\label{eq19}
e^2-e^2_0 =e^2_0 \{\exp [-42q(1+q)Q(t-t_0)]-1\}
\end{equation}
Substituting Equation~(\ref{eq18}) or Equation~(\ref{eq19}) into Equation~(\ref{eq11}), we get
\begin{equation}
\label{eqo24} a=a_0 \exp \left\{ {\frac{19}{7}e_0 ^2\{\exp
[-42q(1+q)Q(t-t_0)]-1\}} \right\},
\end{equation}
\begin{equation}
\label{eqo25} a^2-a_0 ^2=a_0 ^2\left\{ {\exp \left\{
{\frac{38}{7}e_0 ^2\{\exp [-42q(1+q)Q(t-t_0)]-1\}} \right\}-1}
\right\}.
\end{equation}
Substituting the formula~(\ref{eq18}) into the formula~(\ref{eq15}) or the formula~(\ref{eqo25})
into the formula~(\ref{eq13}), we obtain
\begin{equation}
\label{eq20} \Omega =\Omega _0 \exp \left\{
{-\frac{9}{57}\frac{q}{(1+q)}\frac{M}{I}a_0 ^2\left\{ {\exp \left\{
{\frac{38}{7}e_0 ^2\{\exp [-42q(1+q)Q(t-t_0)]-1\}} \right\}-1}
\right\}} \right\}.
\end{equation}

The integrations of the Equations~(\ref{eq7}) -- (\ref{eq9}) can be obtained
\begin{equation}
\label{eq21} \omega =\omega_0 \exp [-\frac{57}{14}(e^2-e^2_0)],
\end{equation}
\begin{equation}
\label{eq22} P_{\rm orb} =(P_0)_{\rm orb} \exp \left\{
{+\frac{57}{14}e_0 \left\{ {\exp [-42q(1+q)Q(t-t_0)]-1} \right\}}
\right\},
\end{equation}
\begin{equation}
\label{eq23} P_{\rm Rot} =(P_{\rm Rot})_0 \exp \left\{
{+\frac{9}{57}\frac{q}{(1+q)}\frac{M}{I}a_0 ^2\left\{ {\exp \left\{
{\frac{38}{7}e_0 ^2\{\exp [-42q(1+q)Q(t-t_0)]-1\}} \right\}-1}
\right\}} \right\}.
\end{equation}

\section{The orbit-spin evolution of the synchronous binaries (Algol A, B
)}

The eclipsing binary system Algol ($\beta $ Per) at least consists of
three components A, B, C. In fact, there is also a massive unseen fourth
component D (\citealt{Hopkins+1976}). Algol A (primary) is a main sequence star (B$_{ 8}$.V). Algol B (secondary) is a subgaint (g K$_{0}$). The
separation between A and B is very near mutually and the system Algol A, B
is regarded as a synchronous binaries due to the action of the tidal
friction. Based on \cite{Giuricin+etal+1984a} the mean rotational angular
velocity of primary A is $v = 56$~km~s$^{-1}$, the mean orbital angular velocity (the
synchronized velocity), $v_{ k} = 55$~km~s$^{-1}$, and based on \cite{Tan+1985}
gave $v_{\rm sini} =55$~km~s$^{-1}$, $v_{\rm syn}= 55$~km~s$^{-1}$. So A and B is a near synchronous
binaries and based on apparent descriptive method for judging the
synchronization of rotation of binary stars (\citealt{Li+2004, Li+2009}) Algol A, B is
a near synchronization binaries. Hence this paper chooses Algol A, B as
synchronous binaries to calculate the orbital circularization and evlution
of the orbit and spin before circularization on the main sequence. For the
data of Algol A, B, we cite the orbital period ($P_{0}$)$_{\rm orb}= 2.8672$~day, $a_0  = 14.03~R_{\odot }$, $M= 3.7~M_{\odot }$,
 $M^{\mbox{'}}=0.81~M_{\odot }$, $R = 2.74~R_{\odot }$, $R^{\mbox{'}}=
3.60~R_{\odot }$, $q={M}'/M=0.22$, $T_{e} = 12010^{ 0}$~K (\citealt{Brancewicz+Dworak+1980}). $e_0 =0.015$ (\citealt{Tomkin+Lambert+1978}, \citealt{Harrington+1984}), $\Omega
_{ 0} = v/R = 2.9365\times 10^{-5}$~rad~s$^{-1}$ and $t_f =t_{f\odot}(M/ M_{\odot })^{ 1/3}(T_{e}/_{e\odot})^{-4/3}$, $t_{f\odot }=0.433$~yr, $T_{e\odot }=5770^{0}$~K, $t_f =0.2519$~yr, $L = 2.2~L_{\odot }$ (\citealt{Popper+1980}).

Where $T_e $ is the effective temperature (\citealt{Zahn+Bouchet+1989b})

Recently, \cite{Yang+etal+2011} present XMM-Newton observation of the
eclipsing binary Algol. Their results of research are valuable.

As \cite{Zahn+Bouchet+1989b} showed that when the coefficient $\lambda ^{lm}$are
all equal, then
\begin{equation}
\lambda =k_2~~~ ({\rm the~apsidal~motion~constant}).
\end{equation}
 $\lambda  = k_{ 2}$ (the apsidal motiom constant) is calculated from
the formula given by \cite{Cowling+1938} and letting $k_{1} = k_{2} = k$
\begin{equation}
\label{eq24}
k=\frac{P_{\rm orb}
/{P}'}{(\frac{R}{a})^5(1+16\frac{M'}{M})+(\frac{R'}{a})^5(1+16(\frac{M}{M'})}
.
\end{equation}
Here $P'$ denotes the period of advance of the apsital line, $P'= 2.476$~year for
Algol A, B given by \cite{Hegedus+1988}. Substituting $P_{\rm orb}$, $P'$, $a$,
$M$, $M'$, $R$ and $R'$ into the above formula, we get
\begin{equation}
\lambda = k = 0.003308.
\end{equation}
The moment of inertia $I$
\begin{equation*}
I = K MR^{2}.
\end{equation*}
$K$ is culculated from the formula
$\frac{1}{K}=\frac{3}{2}(n+\frac{5}{2})$ (\citealt{Schatzman+1963}), The polytropic
index $n = 3$ for the main sequence star (Algol A). So $K = 4/33= 0.1212$.
\begin{equation*}
\frac{M}{I}=\frac{M}{KMR^2}=\frac{1}{KR^2}=0.02268\times 10^{-10}\rm~km^{-2},
~~\exp(152e_{0}^{2}/7)=1.00489.
\end{equation*}
Substituting the values of $k$, $t_f $, $R$, $a$ and $\exp(152e_{0}^{2}/7)$ into
the formula~(\ref{eqo15}), we obtain
\begin{equation}
Q = 2.7 \times  10^{ -8}~\rm yr^{-1}.
\end{equation}
Let us estimate the numerical solutions when the orbit of Algol A, B achieves
the circularizatuion

At first, we evaluate the circularization time scale Substituting the values
of $q$, $R_{0}$, $a_0$, $t_f$ and $\lambda = k = 0.003308$ into the formula~(\ref{eqo10}), we obtain the circularization time scale
\begin{equation}
\label{eq25}
t_{\rm cir} =6.4589\times 10^6\rm~yr.
\end{equation}
Next we estimate the numerical solution of evolutionary tendency of the
orbit-spin when Algol A, B achieved the orbital circularization time. By
letting the intial time $t_{0}= 0$, and substituting $t _{\rm cir}= 6.4589$~yr
into the formulas~(\ref{eq18}), (\ref{eqo24}) or (\ref{eq13})--(\ref{eq23}), we get
$a=14.0226R_{\odot }$, $e=0,0056$, $\omega =2.1931$~rad~d$^{-1}$, $P_{\rm orb}
=2.8649$~d, $\Omega =2.5536$~rad~d$^{-1}$, $P_{\rm Rot} =2.4605$~d; $\delta
a=-0.0074R_{\odot }$, $\delta e=-0.0094$, $\delta \omega =+0.0017$~rad~d$^{-1}$, $\delta P_{\rm orb} =-0.0023$~d, $\delta \Omega =+0.0156$~rad~d$^{-1}$, $\delta P_{\rm Rot} =-0.0160$~d.

The life time (age) is based on stellar mass-lose $\dot {M}$, i.e
\begin{equation}
\label{eq26}
t_{\rm life} =\frac{M}{\dot {M}}.
\end{equation}
The value of its may be calculated from the formule given by \cite{Bowers+Deeming+1984}
\begin{equation}
\label{eq27}
\frac{d(M/M_\Theta)}{dt}=3\times 10^{-8}\frac{(R/R_\Theta)(L/L_\Theta
)}{(M/M_\Theta)}(M_\Theta\rm /yr)
\end{equation}
or calculated from the formula given by \cite{Nieuwenhuijzen+deJager+1990}
\begin{equation}
\label{eq28}
\log\dot {M}=-14.02+1.24\log (L/L_\Theta)+0.81\log (R/R_\Theta)+0.16\log
(M/M_\Theta)
\end{equation}
Substituting these datd of $M$, $R$ and $L$ into the formula~(\ref{eq27}) and (\ref{eq26}), we get
the life time (age)
\begin{equation}
t_{\rm life} =7.5703\times 10^7\rm~yr~(age).
\end{equation}
The lime time for the speed up of spin
\begin{equation}
\label{eq29}
t_\Omega =\frac{\Omega }{\dot {\Omega }}=t_f
/36q^2(\frac{M}{I})ke^2a^2(\frac{R}{a)})^8=4.2447\times 10^8\rm~yr.
\end{equation}
The orbital spiral time (the collasped time of system)
\begin{equation}
\label{eq30} t_a =\frac{a}{\dot {a}}=t_f
/114q(1+q)ke^2(R/a)^8=5.2273\times 10^9\rm~yr.
\end{equation}
\section{Dicussion and conclusions}
\begin{itemize}
\item[(1)] The set of the Equations~(\ref{eq4})--(\ref{eq6}) fits in synchronous binaries
on the pre-main sequence, main sequence and post-main sequence phase
according to the radii of a stars. The raduis of the late type star is
variable due to the gravitational contraction on the pre-main sequence
phase. The raduis of gaint star is variable possibly due to the expansion of
shell on the post-main sequence phase. On the main sequence phase the raduis
of star is stable. Its raduis can be regarded as constant. These refer to
the raduis of primary star because in the Equations~(\ref{eq4})--(\ref{eq6}) $R$ denotes the
raduis of the primary star. It does not refer to the gaint star.

\item[(2)] The research of this paper differs from that of \cite{Zahn+Bouchet+1989b} in some aspects. Zahn \& Boucher's paper researched the
orbital evolution and circularization of binary stars on pre-main sequence
phase by analytical method and for non-synchronized equations by the method
of numerical integration. In the analytical method the radii of binaries are
variable due to gravitational contraction, but the semi-major axis is not
variable including the main sequence phase. However the present paper
studies the orbit-spin of binary stars on the main sequence phase by
analytical method in which the star's raduis is not variable and the
semi-major axis is variable due to tidal friction. In Zahn \& Boucher's
paper they must use the method of numerical integration to solve
non-synchronized equations. However in the present paper the author may use
the analytical method to solve the synchronized equations. Zahn \& Boucher
estimate that the eccentricity decreases from 0.005 to 0.0043 in 10 billion
for binaries 0.5~$M_{\odot }$ + 0.5~$M_{\odot }$ on the main sequence phase.
The present paper estimates that the eccentricity decreases from 0.015 to
0.0056 in 6.45 mega for binaries 3.07~$M_{\odot }+0.81~M_{\odot }$ on the
main sequence phase. Hence both results are different.

\item[(3)] The results of the solution for integrating diferential equations by
analytical mathod are some less different with that by the method of
numerical integration. For example, the semi-major axis
$a=14.0226R_{\odot }$ for Algol A, B when the circularization time
($6.4589\times 10^6$~yr) by the former method and $a=14.0126R_{\odot }$ by
the latter method. However this difference is very small.

\item[(4)] The circularization time occurs before the life time (age), and the lime
time for the speed up of spin and the spiral time (the collasped time of the
system) occur after the life time (age). Hence the latter both are
meaningless.

\item[(5)] In the system of Algol A, B, C the tidal friction in triple stars (\citealt{Kiseleva+etal+1998}) and the perturbing effect of the third star (Algol C)
(\citealt{Li+2006}) may decircularize the orbit of the second star (Algol B). In
this paper one does not consider these effect.
\end{itemize}
One obtained the following conclusions:
\begin{itemize}
\item[(1)] The eccentricity decreases gradually with time until the orbital
circularization, i.e until $e$ decreases to the circularization time.

\item[(2)] The semi-major axis shrinks gradually with time or with eccentricity
decreases.

\item[(3)] The orbital period shortens gradually with time or with circularization.

\item[(4)] The rotational angular velocity of primary component speeds up with time
gradually.

\end{itemize}

\end{document}